# Robust Joint Source-Channel Coding for Delay-Limited Applications

Mahmoud Taherzadeh and Amir. K. Khandani

Coding & Signal Transmission Laboratory(www.cst.uwaterloo.ca)

Dept. of Elec. and Comp. Eng., University of Waterloo, Waterloo, ON, Canada, N2L 3G1

e-mail: {taherzad, khandani}@cst.uwaterloo.ca, Tel: 519-8848552, Fax: 519-8884338

**Abstract**

In this paper, we consider the problem of robust joint source-channel coding over an additive white Gaussian noise channel. We propose a new scheme which achieves the optimal slope of the signal-to-distortion (SDR) curve (unlike the previously known coding schemes). Also, we propose a family of robust codes which together maintain a bounded gap with the optimum SDR curve (in terms of dB). To show the importance of this result, we drive some theoretical bounds on the asymptotic performance of delay-limited hybrid digital-analog (HDA) coding schemes. We show that, unlike the delay-unlimited case, for any family of delay-limited HDA codes, the asymptotic performance loss is unbounded (in terms of dB).

## I. INTRODUCTION

In many applications, delay-limited transmission of analog sources over an additive white Gaussian noise channel is needed. Also, in many cases, the exact signal-to-noise-ratio (SNR) is not known at the transmitter, and may vary over a wide range of values. Two examples of this scenario are transmitting an analog source over a quasi-static fading channel and/or multicasting it to different users (with different channel gains).

Without considering the delay limitations, digital codes can theoretically achieve the optimal performance in the Gaussian channel. Indeed, for the ergodic point-to-point channels, Shannon's source-channel coding separation theorem [1] [2] ensures the optimality of separately designing source and channel codes. However, for the case of limited delay, several articles [3] [4] [5] [6] [7] have shown that joint source-channel codes have a better performance as compared to the separately designed source and channel codes (which are called tandem codes). Also, digital coding is very sensitive to the mismatch in the estimation of the channel SNR.

To avoid the saturation effect of digital coding, various analog and hybrid digital-analog schemes are introduced and investigated in the past [8]–[23]. Among them, examples of 1-to-2-dimensional analog maps can be found as early as the works of Shannon [8] and Kotelnikov [9] and different variations of *Shannon-Kotelnikov maps* (which are also called *twisted modulations*) are studied in [10] [11] [19]. Also, in [14] and [15], analog codes based on dynamical systems are proposed. Although these codes can provide asymptotic gains (for high SNR) over simple repetition codes, they suffer from a threshold effect. Indeed, when the SNR becomes less than a certain threshold, the performance of these systems degrades severely. Therefore, design parameters of these methods should be chosen according to the operating SNR, resulting in sensitivity to SNR estimation errors. Also, although the performance of the system is not saturated for the high SNR values (unlike digital codes), the scaling of the end-to-end distortion is far from the theoretical bounds. Theoretical bounds on the robustness of joint source channel coding schemes (for the delay-unlimited case) are presented in [24] and [25].

To achieve better signal-to-distortion (SDR) scaling, a coding scheme is introduced in [26] [27] which uses $B$ repetitions of a ($k$,$n$) binary code to map the digits of the infinite binary expansion of $k$ samples of the source to the digits of a $nB$-dimensional transmit vector. For this scheme, the bandwidth expansion factor is $\eta = \frac{nB}{k}$ and the SDR asymptotically scales as $SDR \propto SNR^B$, while in theory, the optimum scaling is $SDR \propto SNR^\eta$. Thus, this scheme cannot achieve the optimum scaling by using a single mapping.

In this paper, we address the problem of robust joint source-channel coding, using delay-limited codes. In particular, we show that the optimum slope of the SDR curve can be achieved by a single mapping. The rest of the paper is organized as follows:

In section II, the system model and the basic concepts are presented. Section III presents an analysis of the previous analog coding schemes, and their limitations. In section IV, we introduce a class of joint source-channel codes which have a self-similar structure, and achieve a better asymptotic performance, compared to the other minimum-delay analog and hybrid digital-analog coding schemes. The asymptotic performance of these codes, in terms of the SDR scaling, is comparable with the scheme presented in [26], but with a simpler structure and a shorter delay. We investigate the limits of the asymptotic performance of self-similar coding schemes and their relation with the Hausdorff dimension of the modulation signal set. In section V, we present a single mapping which achieves the optimum slope of the SDR curve, which is equal to the bandwidth expansion factor. Although this mapping achieves the optimum slope of the SDR curve, its gap with the optimum SDR curve is unbounded (in terms of dB). In section VI, we construct a family of robust mappings, which individually achieve the optimum SDR slope, and together, maintain a bounded gap with the optimum SDR curve. We also analyze the limits on the asymptotic performance of the delay-limited HDA coding schemes.

## II. SYSTEM MODEL AND THEORETICAL LIMITS

We consider a memoryless $\{X_i\}_{i=1}^{\infty}$ uniform source with zero mean and variance $\frac{1}{12}$, i.e. $-\frac{1}{2} \leq x_i < \frac{1}{2}$. Also, the samples of the source sequence are assumed independent with identical distributions (i.i.d.). Although the focus of this paper is on a source with uniform distribution, as it is discussed in Appendix C, the asymptotic results are valid for all distributions which have a bounded probability density function.

The transmitted signal is sent over an additive white Gaussian noise (AWGN) channel. The problem is to map the one-dimensional signal to the $N$-dimensional channel space, such that the effect of the noise is minimized. This means that the data $x$, $-\frac{1}{2} \leq x < \frac{1}{2}$, is mapped to the transmitted vector $\mathbf{s} = (s_1, ..., s_N)$. At the receiver side, the received signal is $\mathbf{y} = \mathbf{s} + \mathbf{z}$ where $\mathbf{z} = (z_1, ..., z_N)$ is the additive white Gaussian noise with variance $\sigma^2$.

As an upper bound on the performance of the system, we can consider the case of delay-unlimited. In this case, we can use Shannon's theorem on the separation of source and channel coding. By combining the lower bound on the distortion of the quantized signal (using the rate-distortion formula) and the capacity of $N$ parallel Gaussian channels with the noise variance $\sigma^2$, we can bound the distortion $D = \mathrm{E}\{|x - \widetilde{x}|^2\}$ as [15]

$$D \geq c\sigma^{2N} \tag{1}$$

where $c$ is a constant number.

## III. CODES BASED ON DYNAMICAL SYSTEMS AND HYBRID DIGITAL-ANALOG CODING

Previously, two related schemes, based on dynamical systems, have been proposed for the scenario of delay-limited analog coding:

1) Shift-map dynamical system [14]
2) Spherical shift-map dynamical system [15]

These are further explained in the following.

### A. Shift-map dynamical system

In [14], an analog transmission scheme based on shift-map dynamical systems is presented. In this method, the analog data $x$ is mapped to the modulated vector $(s_1, ..., s_N)$ where

$$s_1 = x \mod 1 \tag{2}$$

$$s_{i+1} = b_i s_i \mod 1, \quad \text{for } 1 \leq i \leq N-1 \tag{3}$$

where $b_i$ is an integer number, $b_i \geq 2$. The set of modulated signals generated by the shift map consists of $b_1 \cdot b_2 \cdot ... \cdot b_{N-1}$ parallel segments inside an $N$-dimensional unit hypercube. In [15], the authors have shown that by appropriately choosing the parameters $\{b_i\}$ for different SNR values, one can achieve the SDR scaling (versus the channel SNR) with the slope $N - \epsilon$, for any positive number $\epsilon$. Indeed, we can have a slightly tighter upper bound on the end-to-end distortion as follows:

**Theorem 1** *Consider the shift-map analog coding system which maps the source sample to an $N$-dimensional modulated vector. For any noise variance[1] $\sigma^2 \leq \frac{1}{2}$, we can find parameter $a$ such that for the shift-map scheme with the parameters $b_i = a \geq 2$, the distortion of the decoded signal $D$ is bounded as[2]*

$$D \leq c\sigma^{2N}(-\log \sigma)^{N-1} \tag{4}$$

*where $c$ depends only on $N$.*

---

[1] The result is still valid if $\sigma^2 \leq \delta$, for some $0 < \delta < 1$ (but $c$ will depend on $\delta$).
[2] We use $\log x$ to denote the natural logarithm, i.e. $\log_e x$.

*Proof:* See Appendix A. ∎

Also, we have the following lower bound on the end-to-end distortion:

**Theorem 2** *For any shift-map analog coding scheme and any noise variance $\sigma^2 \leq \frac{1}{2}$, the output distortion is lower bounded as*

$$D \geq c'\sigma^{2N}(-\log \sigma)^{N-1} \tag{5}$$

*where $c'$ depends only on $N$.*

*Proof:* See Appendix B. ∎

## B. Spherical shift-map dynamical system

In [15], a spherical code based on the linear system $\dot{\mathbf{s}}_T = \mathbf{A}\mathbf{s}_T$ is introduced, where $\mathbf{s}_T$ is the $2N$-dimensional modulated signal and $\mathbf{A}$ is a skew-symmetric matrix, i.e. $\mathbf{A}^T = -\mathbf{A}$. This scheme is very similar to the shift-map scheme. Indeed, with an appropriate change of coordinates, the above modulated signal can be represented as

$$\mathbf{s}_T = \frac{1}{\sqrt{N}} \Big(\cos 2\pi x, \cos 2a\pi x, ..., \cos 2a^{N-1}\pi x,$$

$$\sin 2\pi x, \sin 2a\pi x, ..., \sin 2a^{N-1}\pi x\Big) \tag{6}$$

for some parameter $a$.

If we consider $\mathbf{s}_{sm}$ as the modulated signal generated by the shift-map scheme with parameters $b_i = a$ in (3), then, (6) can be written in the vector form as

$$\mathbf{s}_T = \left(Re\left\{e^{\pi i \mathbf{s}_{sm}}\right\}, Im\left\{e^{\pi i \mathbf{s}_{sm}}\right\}\right). \tag{7}$$

The relation between the spherical code and the linear shift-map code is very similar to the relation between phase-shift-keying (PSK) and pulse-amplitude-modulation (PAM). Indeed, the spherical shift-map code and PSK modulation are, respectively, the linear shift-map and PAM modulations which are transformed from the unit interval, $[\frac{-1}{2}, \frac{1}{2})$, to the unit circle.

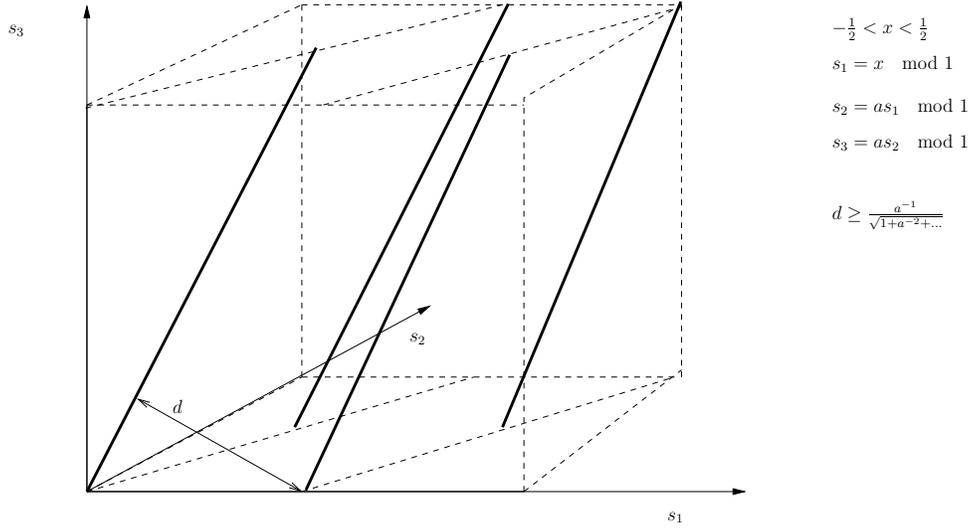

Fig. 1. The shift-map modulated signal set for $N = 3$ dimensions and $a = 2$.

For the performance of the spherical codes, the same result as Theorem 1 is valid. Indeed, for any parameters $a$ and $N$, the spherical code asymptotically has a saving of $\frac{(2\pi)^2}{12}$ or 5.17 dB in the power. This asymptotic gain results from transforming the unit-interval signal set (with length 1 and power $\frac{1}{12}$) to the unit-circle signal set (with length $2\pi$ and power 1) . However, the spherical code uses $2N$ dimensions (compared to $N$ dimensions for the linear shift-map scheme).

For both these methods, for any fixed parameter $a$, the output SDR asymptotically has linear scaling with the channel SNR. The asymptotic gain (over the simple repetition code) is approximately $a^{2(N-1)}$ (because the modulated signal is stretched approximately $a^{N-1}$ times)[3]. Therefore, a larger scaling parameter $a$ results in a higher asymptotic gain. However, by increasing $a$, the distance between the parallel segments of the modulated signal set decreases. This distance is approximately $\frac{1}{a}$ and for the low SNRs (when the noise variance is larger than or comparable to $\frac{1}{a}$), jumping from one segment of the modulated signal set to another one becomes the dominant factor in the distortion of the decoded signal which results in a poor performance in this SNR region. Thus, there is a trade-off between the gain in the high-SNR region and the critical noise level which is fatal for the system. By increasing the scaling parameter $a$, the asymptotic gain increases, but at the same time, a higher SNR threshold is needed to achieve that gain. In [28], the authors have combined the dynamical-system schemes with LDPC and iterative decoding to reduce the critical SNR threshold. However, overall behavior of the output distortion is the same for all these methods. Also, in [29] and [30], a scheme is introduced for approaching arbitrarily close to the optimum SDR, for colored sources. However, it is not delay-limited and it only works for the bandwidth expansion of 1.

---

[3]The exact asymptotic gain is equal to the scaling factor of the signal set, i.e. $a^{2(N-1)}\left(1 + \frac{1}{a^2} + ... + \frac{1}{a^{2(N-1)}}\right)$ for the shift map and $\frac{(2\pi)^2}{12}a^{2(N-1)}\left(1 + \frac{1}{a^2} + ... + \frac{1}{a^{2(N-1)}}\right)$ for the spherical shift map.

The shift-map analog coding system can be seen as a variation of a hybrid-digital-analog (HDA) joint source-channel code. Various types of such hybrid schemes are investigated in [16] [17] [18] [24] and [31]. Indeed, for the shift-map system, we can rotate the modulated signal set such that all the parallel segments of it become aligned in the direction of one of the dimensions. In this case, by changing the support region of the modulated set (which is a rotated $N$-dimensional cube) to the standard cube, we obtain a new similar modulation which is hybrid digital-analog and has almost the same performance. In the new modulation, the information signal is quantized by $a^{N-1}$ points in an $(N-1)$-dimensional sub-space and the quantization error is transmitted over the remaining dimension.

Regarding the scaling of the output distortion, the performance of the shift-map scheme, with appropriate choice of parameters for each SNR, is very close to the theoretical limit. In fact, the output distortion scales as $\sigma^{2N}(-\log \sigma)^{N-1}$, instead of being proportional to $\sigma^{2N}$. However, for any fixed set of parameters, the curve of SDR-versus-SNR (in dB) is saturated by the unit slope (instead of $N$). This shortcoming is an inherent drawback of schemes like the shift-map code or the spherical code (which are based on dynamical systems). Indeed, in [32], it is shown that no single differentiable mapping can achieve an asymptotic slope better than 1. This article addresses this shortcoming.

There are some other analog codes in the literature which use different mappings. Analog codes based on the 2-dimensional Shannon map [20] [21] [22] [23], or the tent map [14] are examples of these codes. However, all these codes share the shortcomings of the shift-map code.

## IV. JOINT SOURCE-CHANNEL CODES BASED ON FRACTAL SETS

In this section, we propose a coding scheme, based on the fractal sets, that can achieve slopes greater than 1 (for the curve of SDR versus SNR).

**Scheme I:** For the modulating signal $x$, $-\frac{1}{2} \leq x < \frac{1}{2}$, we consider the binary expansion of $x + \frac{1}{2}$:

$$x + \frac{1}{2} = \left(\overline{0 \cdot b_1 b_2 b_3 ...}\right)_2. \tag{8}$$

Now, we construct $s_1, s_2, ..., s_N$ as

$$s_1 = \left(\overline{0 \cdot b_1 b_{N+1} b_{2N+1} ...}\right)_\alpha \tag{9}$$

$$s_2 = \left(\overline{0 \cdot b_2 b_{N+2} b_{2N+2} ...}\right)_\alpha \tag{10}$$

$$...$$

$$s_N = \left(\overline{0 \cdot b_N b_{2N} b_{3N} ...}\right)_\alpha \tag{11}$$

where $\left(\overline{0 \cdot b_1 b_2 b_3 ...}\right)_\alpha$ is the base-$\alpha$ expansion[4].

**Theorem 3** *In the proposed scheme, for any $\alpha > 2$ and noise variance $\sigma^2 \leq \frac{1}{2}$, the output distortion $D$ is upper bounded by*

$$D \leq c \sigma^{2\beta} (-\log \sigma)^N \tag{12}$$

*where $c$ depends only on $N$, and $\beta = N \frac{\log 2}{\log \alpha}$.*

*Proof:* Consider $z_i$ as the Gaussian noise on the $i$th dimension:

$$\Pr\left\{|z_i| > 2\sqrt{N}\sigma\sqrt{-\log \sigma}\right\} = \tag{13}$$

$$2Q\left(2\sqrt{N}\sqrt{-\log \sigma}\right) \leq e^{-\frac{4N(-\log \sigma)}{2}} = e^{-2N(-\log \sigma)} = \sigma^{2N} \tag{14}$$

---

[4]In this article, we define the base-$\alpha$ expansion, for any real number $\alpha > 2$ and any binary sequence $(b_1 b_2 b_3 ...)$, as $\left(\overline{0 \cdot b_1 b_2 b_3 ...}\right)_\alpha \triangleq \sum_{i=1}^{\infty} b_i \alpha^{-i}$.

Now, we bound the distortion, conditioned on $|z_i| \leq 2\sqrt{N}\sigma\sqrt{-\log \sigma}$ for $1 \leq i \leq N$. If the $k$th digit of $s_i$ and $s'_i$ are different,

$$|s_i - s'_i| \geq \tag{15}$$

$$\left(\overline{0 \cdot \underbrace{0...0}_{k-1}1000..}\right)_\alpha - \left(\overline{0 \cdot \underbrace{0...0}_{k-1}0111...}\right)_\alpha \tag{16}$$

$$> (\alpha - 2)\alpha^{-(k+1)} \tag{17}$$

Therefore, if $|s_i - s'_i| \leq \delta$ for $\delta > 0$, the first $k$ digits of $s_i$ and $s'_i$ are the same, where $k \geq \lfloor -\log_\alpha\left(\frac{\delta}{\alpha-2}\right)\rfloor - 1$. Now, by considering $\delta = 4\sqrt{N}\sigma\sqrt{-\log \sigma}$,

$$|s_i - s'_i| \leq 2|z_i| \leq 4\sqrt{N}\sigma\sqrt{-\log \sigma} \tag{18}$$

$$\implies k \geq \left\lfloor -\log_\alpha\left(\frac{4\sqrt{N}\sigma\sqrt{-\log \sigma}}{\alpha - 2}\right)\right\rfloor - 1 \tag{19}$$

Therefore, for $1 \leq i \leq N$, the first

$$\left\lfloor -\log_\alpha\left(\frac{4\sqrt{N}\sigma\sqrt{-\log \sigma}}{\alpha - 2}\right)\right\rfloor - 1$$

digits of $s_1, s_2, ..., s_N$ can be decoded without any error, hence, the first

$$N\left(\left\lfloor -\log_\alpha\left(\frac{4\sqrt{N}\sigma\sqrt{-\log \sigma}}{\alpha - 2}\right)\right\rfloor - 1\right)$$

bits of the binary expansion of $x$ can be reconstructed perfectly. In this case, the output distortion is bounded by

$$\sqrt{D} \leq 2^{-N\left(\left\lfloor -\log_\alpha\left(\frac{4\sqrt{N}\sigma\sqrt{-\log \sigma}}{\alpha-2}\right)\right\rfloor - 1\right)} \tag{20}$$

$$\implies D \leq c_1 \sigma^{2\beta}(-\log \sigma)^N \tag{21}$$

where $c_1$ depends only on $\alpha$ and $N$. By combining the upper bounds for the two cases, noting that $\sigma < 1$

$$D \leq \sum_{i=1}^{N} \Pr\left\{|z_i| > 2\sqrt{N}\sigma\sqrt{-\log \sigma}\right\} + c_1 \sigma^{2\beta}(-\log \sigma)^N \tag{22}$$

$$\leq \sigma^{2N} + c_1 \sigma^{2\beta}(-\log \sigma)^N \tag{23}$$

$$\leq c\sigma^{2\beta}(-\log \sigma)^N. \tag{24}$$

∎

According to the theorem 2, for any $\epsilon > 0$, we can construct a modulation scheme that achieves the asymptotic slope of $N - \epsilon$ (for the curve of SDR versus SNR, in terms of dB). As expected (according to the result by Ziv [32]), none of these mappings are differentiable. More generally, Ziv has shown that [32]:

**Theorem 4** *( [32], Theorem 2) For the modulation mapping* $\mathbf{s} = f(x)$, *define*

$$d_f(\Delta) = \mathrm{E}\left\{\|f(x + \Delta) - f(x)\|^2\right\}.$$

*If there are positive numbers* $A, \gamma, \Delta_0$ *such that*

$$d_f(\Delta) \leq A\Delta^\gamma \text{ for } \Delta \leq \Delta_0. \tag{25}$$

*Then, there is constant $c$ such that*

$$D \geq c\sigma^{\frac{2}{\gamma}}. \tag{26}$$

In Scheme I, by decreasing $\alpha$, we can increase the asymptotic slope $\beta$. However, it also degrades the low-SNR performance of the system. This phenomenon is observed in figure 3.

In scheme I, the signal set is a self-similar fractal [33], where the parameter $\beta$, which determines the asymptotic slope of the curve, is the dimension of the fractal. There are different ways to define the fractal dimension. One of them is the Hausdorff dimension. Consider $\mathcal{F}$ as a Borel set in a metric space, and $\mathfrak{A}$ as a countable family of sets that covers it. We define $H^s_\varepsilon(\mathcal{F}) = \inf \sum_{\mathcal{A} \in \mathfrak{A}} (\mathrm{diameter}(\mathcal{A}))^s$, where the infimum is over all countable covers that diameter of their sets are not larger than $\varepsilon$. The $s$-dimensional Hausdorff space is defined as $H^s(\mathcal{F}) = \lim_{\varepsilon \to 0} H^s_\varepsilon(\mathcal{F}) = \sup_{\varepsilon > 0} H^s_\varepsilon(\mathcal{F})$. It can be shown that there is a critical value $s_0$ such that for $s < s_0$, this measure is infinite and for $s > s_0$, it is zero [33]. This critical value $s_0$ is called the Hausdorff dimension of the set $\mathcal{F}$.

Another useful definition is the box-counting dimension. If we partition the space into a grid of cubic boxes of size $\varepsilon$, and consider $m_\varepsilon$ as the number of boxes which intersect the set $\mathcal{F}$, the box-counting dimension of $\mathcal{F}$ is defined as

$$\text{Dim}_b(\mathcal{F}) = \lim_{\varepsilon \to 0} \frac{\log m_\varepsilon}{\log \frac{1}{\varepsilon}} \qquad (27)$$

It can be shown that for regular self-similar fractals, the Hausdorff dimension is equal to the box-counting dimension [33]. Intuitively, theorem 3 means that in scheme I among the $N$ available dimensions, only $\beta$ dimensions are effectively used. Indeed, we can show that for any modulation set[5] with box-counting dimension $\beta$, the asymptotic slope of the SDR curve is at most $\beta$:

**Theorem 5** *For a modulation mapping* $\mathbf{s} = f(x)$, *if the modulation set $\mathcal{F}$ has box-counting dimension $\beta$, then*

$$\lim_{\sigma \to 0} \frac{\log D}{\log \sigma} \leq 2\beta. \qquad (28)$$

*Proof:* We divide the space to boxes of size $\sigma$. Consider $m_\sigma$ as the number of cubic boxes that cover $\mathcal{F}$. We divide the source signal set to $4m_\sigma$ segments of length $\frac{1}{4m_\sigma}$. Consider $\mathcal{A}_1, ..., \mathcal{A}_{4m_\sigma}$ as the corresponding $N$-dimensional optimal decoding regions (based on the MMSE criterion), and $\mathcal{B}_1, ..., \mathcal{B}_{4m_\sigma}$ as their intersection with the $m_\sigma$ cubes (see figure 2). Total volume of these $4m_\sigma$ sets is equal to the total volume of the covering boxes, i.e. $m_\sigma \sigma^N$. Thus, at least, half of these sets (i.e. $2m_\sigma$ of them) have volume less than $\frac{1}{2}\sigma^N$. For any of these sets such as $\mathcal{B}_i$ and any box, the volume of the intersection of that box with the other sets is at least $V_{min} = \sigma^N - \frac{1}{2}\sigma^N = \frac{1}{2}\sigma^N$. For any point in the corresponding segments of the set $\mathcal{B}_i$, the probability of decoding to a wrong segment is lower bounded by the probability of a jump to the neighboring sets in the same box. Because the variance of the additive Gaussian noise is $\sigma^2$ per each dimension, and for such a jump the squared norm of the noise at most needs to be $N\sigma^2$ (square of the diameter of the box), the probability of such a jump to the neighboring sets can be lower bounded as

$$\Pr(jump) \geq V_{min} \cdot \min_{\|\mathbf{z}\|^2 \leq N\sigma^2} f_{\mathbf{z}}(\mathbf{z}) \quad (29)$$

$$\geq \frac{1}{2}\sigma^N \cdot \frac{1}{(2\pi)^{\frac{N}{2}}\sigma^N} e^{-\frac{N\sigma^2}{2\sigma^2}} = \frac{1}{2^{\frac{N}{2}+1}\pi^{\frac{N}{2}}} e^{-\frac{N}{2}}, \quad (30)$$

where $f_{\mathbf{z}}(\mathbf{z})$ is the pdf of the noise vector $\mathbf{z}$.

---
[5]Modulation set is the set all possible modulated vectors.

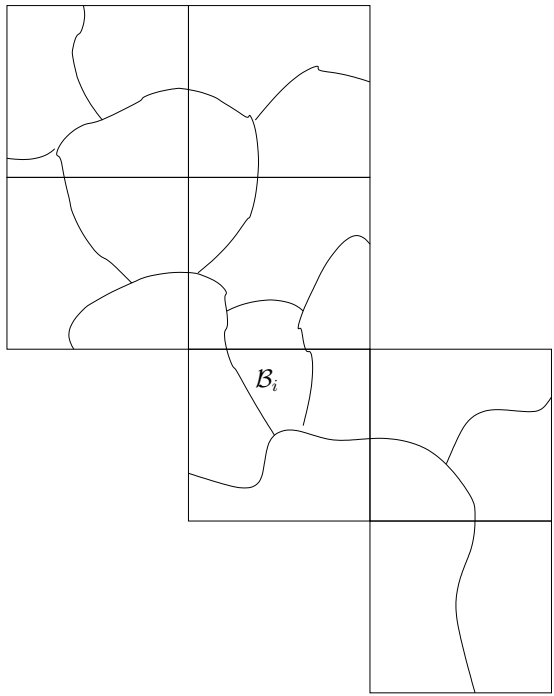

Fig. 2. Boxes of size $\sigma$ and their intersections with the decoding regions

Now, for these segments of the source, consider the subsegments with length $\frac{1}{20m_\sigma}$ at the center of them. When the source belongs to one of these subsegments, wrong segment decoding results in a squared error of at least $\left(\frac{1}{2} \cdot \left(\frac{1}{4m_\sigma} - \frac{1}{20m_\sigma}\right)\right)^2 = \left(\frac{1}{10m_\sigma}\right)^2$. Thus, for these subsegments whose total length is at least $\frac{1}{20m_\sigma} \cdot 2m_\sigma = \frac{1}{10}$, at least with probability $\Pr(jump)$, we have a squared error which is not less than $\left(\frac{1}{10m_\sigma}\right)^2$. Therefore,

$$D \geq \frac{1}{10}\Pr(jump) \cdot \left(\frac{1}{10m_\sigma}\right)^2 = \frac{c}{m_\sigma^2} \quad (31)$$

where $c$ only depends on the bandwidth expansion $N$. On the other hand, based on the definition of the box-counting dimension,

$$\beta = \lim_{\sigma \to 0} \frac{\log m_\sigma}{\log \frac{1}{\sigma}}. \tag{32}$$

By using (31) and (32),

$$\lim_{\sigma \to 0} \frac{\log D}{\log \sigma} \leq 2\beta. \tag{33}$$

∎

It should be noted that theorem 5 is valid for all signal sets, not just self-similar signal sets. As a corollary, based on the fact that the box-counting dimension can not be greater than the dimension of the space [33], Theorem 5 provides a geometric insight to (1).

Another scheme based on self-similar signal sets and the infinite binary expansion of the source is proposed in [26] [27], which similar to the scheme proposed in this section, achieves a SDR scaling better than linear coding, but cannot achieve the optimum SDR scaling. The scheme presented in [26] is based on using $B$ repetitions of a ($k$,$n$) binary code to map the digits of the infinite binary expansion of $k$ samples of the source to the digits of a $nB$-dimensional transmit vector. This scheme shares the shortcoming of Scheme I. In [26], the bandwidth expansion factor is $\eta = \frac{nB}{k}$ and the SDR asymptotically scales as $SDR \propto SNR^B$, instead of the optimum scaling $SDR \propto SNR^\eta$. The main difference between Scheme I and the scheme proposed in [26] is that in Scheme I, the delay is minimum (it uses only one sample of the source for coding), but in [26], the delay is $k$, and the the ratio between the SDR exponent and the optimum SDR exponent is dependent on the delay (it is $\frac{k}{n}$), i.e. to increase it, one needs to increase the length of the binary code, which results in increasing the delay.

The idea of using the infinite binary expansion of the source, for joint source-channel coding, can be traced back to Shannon's 1949 paper [8], where shuffling the digits is proposed for bandwidth contraction (i.e. mapping high-dimensional data to a signal set with a lower dimension). For bandwidth expansion, space-filling self-similar signal sets have been investigated in [13], however, the SDR scaling of those schemes are not better than linear coding. The reason is that when we use a self-similar set to fill the space, the squared error caused by jumping to adjacent subsets dominates the scaling of the distortion. To avoid this effect, we need to avoid filling the whole space. This results in losing dimensionality for self-similar sets, which results in sub-optimum SDR scaling (as investigated in this section). To avoid this drawback, we need to consider signal sets which are not self-similar, as proposed in the next section.

## V. ACHIEVING THE OPTIMUM ASYMPTOTIC SDR SLOPE USING A SINGLE MAPPING

Although Scheme I can construct mappings that achieve near-optimum slope for the curve of SDR (versus the channel SNR), none of these mappings can achieve the optimum slope $N$. To achieve the optimum slope with a single mapping, we slightly modify Scheme I:

For the modulating signal $x$, consider $x + \frac{1}{2} = \left(\overline{0.b_1 b_2 b_3 ...}\right)_2$. We construct $s_1, s_2, ..., s_N$ as

$$s_1 = \left(\overline{0.b_1 0 b_{\frac{N(N+1)}{2}+1} b_{\frac{N(N+1)}{2}+2} ... b_{\frac{N(N+1)}{2}+N+1} 0 b_{\frac{(2N)(2N+1)}{2}+1} ...}\right)_2 \tag{34}$$

$$s_2 = \left(\overline{0.b_2 b_3 0 b_{\frac{(N+1)(N+2)}{2}+1} b_{\frac{(N+1)(N+2)}{2}+2} ... b_{\frac{(N+1)(N+2)}{2}+N+2} 0 ...}\right)_2 \tag{35}$$

...

...

$$s_N = \left(\overline{0.b_{\frac{N(N-1)}{2}+1} b_{\frac{N(N-1)}{2}+2} ... b_{\frac{N(N+1)}{2}} 0 ...}\right)_2 \tag{36}$$

The difference between this scheme and Scheme I is that instead of assigning the $kN+i$th bit to the signal $s_i$, the bits of the binary expansion of $x + \frac{1}{2}$ are grouped such that the $l$th group ($l = kN + i$) consists of $l$ bits and is assigned to the $i$th dimension. In decoding, we find the point in the signal set which is closest to the received vector $\mathbf{s} + \mathbf{z}$. If $|z_i| < 2^{-1-\sum_{k=0}^{n}(kN+i+1)}$, the first $\sum_{k=0}^{n}(kN+i+1)$ bits of $s_i$ can be decoded error-freely (for $1 \leq i \leq N$) which include $\sum_{k=0}^{n}(kN+i)$ bits of the source $x$.

**Theorem 6** *Using the mapping constructed by Scheme II, for any noise variance $\sigma^2 \leq \frac{1}{2}$, the output distortion $D$ is upper bounded by*

$$D \leq c_1 \sigma^{2N} 2^{c_2 \sqrt{-\log_2 \sigma}} \tag{37}$$

*where $c_1$ and $c_2$ only depend[6] on $N$.*

---

[6]Throughout this paper, $c_1, c_2, ...$ are constants, independent of $\sigma$ (they may depend on $N$).

*Proof:* Let $z_i$ be the Gaussian noise on the $i$th dimension and assume that $n$ is selected such that

$$\sum_{k=1}^{n+1}(kN+1) \leq -\log_2 \sigma < \sum_{k=1}^{n+2}(kN+1). \qquad (38)$$

The probability that $|z_i| \geq 2^{-1-\sum_{k=1}^{n}(kN+1)}$ is negligible. Indeed,

$$\Pr\left\{|z_i| \geq 2^{-1-\sum_{k=1}^{n}(kN+1)} \,\middle|\, -\log_2 \sigma \geq \sum_{k=1}^{n+1}(kN+1)\right\} \leq \qquad (39)$$

$$2Q\left(\frac{2^{-1-\sum_{k=1}^{n}(kN+1)}}{2^{-\sum_{k=1}^{n+1}(kN+1)}}\right) = 2Q\left(2^{(n+1)N}\right) \qquad (40)$$

$$\stackrel{a}{<} 2Q\left(2^{\sqrt{\sum_{k=1}^{n+2} kN+1}}\right) \qquad (41)$$

$$\stackrel{b}{<} 2Q\left(2^{\sqrt{-\log_2 \sigma}}\right) \stackrel{c}{<} 2^{-2^{2\sqrt{-\log_2 \sigma}-1}} \qquad (42)$$

where $(a)$ because $\sqrt{\sum_{k=1}^{n+2} kN+1} = \sqrt{\frac{N(n+2)(n+3)}{2}+n+2} < (n+1)N$ for $N \geq 2$, and $(b)$ because (38), and $(c)$ because $Q(x) < \frac{1}{2}e^{-\frac{x^2}{2}}$.

On the other hand, when $|z_i| < 2^{-1-\sum_{k=1}^{n}(kN+1)}$, for $1 \leq i \leq N$, $|z_i| < 2^{-1-\sum_{k=0}^{n-1}(kN+i+1)}$, hence the first $\sum_{k=0}^{n-1}(kN+i+1)$ bits of $s_i$ can be decoded error-freely which include $\sum_{k=0}^{n-1}(kN+i)$ bits of the source $x$. Thus, the first $\sum_{i=1}^{N}\sum_{k=0}^{n-1}(kN+i) = \sum_{j=1}^{nN} j$ bits of $x$ can be decoded error-freely. Now,

$$\sum_{j=1}^{nN} j = \frac{nN(nN+1)}{2} \qquad (43)$$

$$= N\left(\frac{N(n+2)(n+3)}{2}+n+2\right) - \frac{N^2(5n+6)+nN+4N}{2} \qquad (44)$$

$$= N\left(\sum_{k=1}^{n+2}(kN+1)\right) - \frac{N^2(5n+6)+nN+4N}{2} \qquad (45)$$

$$\geq N\left(\sum_{k=1}^{n+2}(kN+1)\right) - c_3\sqrt{\sum_{k=1}^{n+1}(kN+1)} \qquad (46)$$

where $c_3$ depends only on $N$. Therefore, by using the assumption (38),

$$\sum_{j=1}^{nN} j \geq \tag{47}$$

$$- N \log_2 \sigma - c_3 \sqrt{-\log_2 \sigma} \tag{48}$$

Consequently, the output distortion is bounded by

$$D \leq \sum_{i=1}^{N} \Pr\left\{|z_i| \geq 2^{-1-\sum_{k=1}^{n} kN}\right\} + 2^{-2\sum_{j=1}^{nN} j} \tag{49}$$

$$\leq 2Q\left(2^{\sqrt{-\log_2 \sigma}}\right) + 2^{2N \log_2 \sigma + 2c_3 \sqrt{-\log_2 \sigma}} \tag{50}$$

$$= 2Q\left(2^{\sqrt{-\log_2 \sigma}}\right) + \sigma^{2N} 2^{c_2 \sqrt{-\log_2 \sigma}}. \tag{51}$$

$$\Longrightarrow D \leq c_1 \sigma^{2N} 2^{c_2 \sqrt{-\log_2 \sigma}}. \tag{52}$$

∎

It should be noted that in this proof, the assumption of having a uniform distribution is not used, and the above proof is valid for any source whose samples are in the interval $\left[-\frac{1}{2}, \frac{1}{2}\right)$. In Appendix C, we extend the scheme proposed in this section to other sources which are not necessarily bounded.

## VI. APPROACHING A NEAR-OPTIMUM SDR BY DELAY-LIMITED CODES

In [24], a family of hybrid digital-analog (HDA) source-channel codes are proposed which together can achieve the optimum SDR curve and each of them only suffers from the mild saturation effect (the asymptotic unit slope for the curve of SDR versus SNR). However, their approach is based on using capacity-approaching digital codes as a component of their scheme. In [25], it is shown that for any joint source-channel code that touches the optimum SDR curve at a certain SNR point, the asymptotic slope can not be better than one.

In this section, we consider the problem of finding a family of delay-limited analog codes which together have a bounded asymptotic loss in the SDR performance (in terms of dB). Results of Section III show that none the previous analog coding schemes (based on dynamical systems) can construct such a family of codes. In this section, we also show that no HDA source-channel coding scheme can achieve this goal.

In the HDA source-channel coding, in general, to map an $M$ dimensional source to an $N$ dimensional signal set, the source is quantized by $\kappa$ points which are sent over $N-M$ dimensions and the residual noise is transmitted over the remaining $M$ dimensions. In other words, the region of the source (which is a hypercube for the case of a uniform source) is divided into $\kappa$ subregions $\mathcal{A}_1, ..., \mathcal{A}_\kappa$. These subregions are mapped to $\kappa$ parallel subsets of the $N$ dimensional Euclidean space, $\mathcal{A}'_1, ..., \mathcal{A}'_\kappa$, where $\mathcal{A}'_i$ is a scaled version of $\mathcal{A}_i$ with a factor of $a$.

**Theorem 7** *Consider a HDA joint source-channel code which maps an $M$-dimensional uniform source (inside the unit cube) to $\kappa$ parallel $M$-dimensional subsets of an $N$ dimensional Euclidean space ($N > M$), with a power constraint of $1$. If the decoding of digital and analog parts are done separately, for any noise variance $\sigma^2 < 1$, the output distortion is lower bounded by*

$$D \geq c\sigma^{\frac{2N}{M}}(-\log \sigma)^{\frac{N-M}{M}} \tag{53}$$

*where $c$ depends only on $M$ and $N$.*

*Proof:* See Appendix D. ∎

Now, we construct families of delay-limited analog codes which by a proper choice of parameters (according to the channel SNR) have a bounded asymptotic loss in the SDR performance (in terms of dB).

**Type I - Family of piece-wise linear mappings:** For any $2^{-k-1} < \sigma \leq 2^{-k}$, for $k > 0$, we construct an analog code as the following:

For $x + \frac{1}{2} = \overline{(0 \cdot b_1 b_2 ... b_{Nk-1})}_2 + \frac{\{2^{Nk-1}x\}}{2^{Nk-1}}$, where $\{\cdot\}$ represents the fractional part, we construct $s_1, s_2, ..., s_N$ as

$$s_1 = \sum_{i=1}^{k}(2^{-i} + 2^{-k}(k-i))b_{(i-1)N+1}$$

$$s_2 = \sum_{i=1}^{k}(2^{-i} + 2^{-k}(k-i))b_{(i-1)N+2}$$

$$\ldots$$

$$s_{N-1} = \sum_{i=1}^{k}(2^{-i} + 2^{-k}(k-i))b_{(i-1)N+N-1}$$

$$s_N = \sum_{i=1}^{k-1}(2^{-i} + 2^{-k}(k-i))b_{(i-1)N+N} + 2^{Nk-k-2}\frac{\{2^{Nk-1}x\}}{2^{Nk-1}} \quad (54)$$

First, we show that $0 \leq s_j < 2$, for $1 \leq j \leq N$. By using the fact that the value of the bits are at most 1, and $\{2^{Nk-1}x\} < 1$,

$$s_j \leq \sum_{i=1}^{k}(2^{-i} + 2^{-k}(k-i)) + 2^{-k-1} = \sum_{i=1}^{k+1} 2^{-i} + 2^{-k}\sum_{i=1}^{k}(k-i) \quad (55)$$

$$< 1 + 2^{-k} \cdot \frac{k(k-1)}{2} < 2. \quad (56)$$

Therefore, noting that $0 \leq s_j < 2$, by an appropriate shift (e.g. modifying the transmitted signal set as $\mathbf{s}' = \mathbf{s} - 1$), the transmitted power can be bounded by one. Next, we show that the proposed scheme has a bounded gap (in terms of dB) to the optimum SDR curve:

**Theorem 8** *In the proposed scheme, noise variance $\sigma^2 \leq \frac{1}{2}$, the output distortion $D$ is upper bounded by*

$$D \leq c\sigma^{2N} \quad (57)$$

*where $c$ depends only on $N$.*

*Proof:* The signal set consists of $2^{Nk-1}$ segments of length $2^{-k-1}$, where each of them is a subsegment of the source region (the unit interval), scaled by a factor of $2^{Nk-k-2}$.

The probability that the first error occurs in the $l$th bit ($l = (i-1)N + j$, where $1 \leq j \leq N$) of $x$ is bounded by $P_l \leq 2Q\left(\frac{k-i}{2}\right) \leq 2Q\left(\frac{k}{2} - \frac{l}{2N}\right)$ and it results in an output squared error of at most $D_l \leq 4^{-l+1} = 4^{-(i-1)N-j+1}$. Therefore, by considering the union-bound over all possible errors, we obtain

$$D \leq \sum_{l=1}^{Nk-1} D_l \cdot P_l + D_{no-bit-error}$$

$$\leq \sum_{l=1}^{Nk-1} 4^{-l+1} \cdot 2Q\left(\frac{k}{2} - \frac{l}{2N}\right) + 4^{-(Nk-k-2)}\sigma^2. \tag{58}$$

Now, by using $Q(x) < e^{-\frac{x^2}{2}}$ and $2^{-k-1} < \sigma$, we have

$$D \leq \sum_{l=1}^{kN-1} 2^{-2l+3} e^{-\frac{(k-l/N)^2}{8}} + 4^{-(Nk-k-2)}\sigma^2$$

$$\leq \sum_{l=1}^{kN-1} 2^{-2l+3} 2^{-\frac{(k-l/N)^2}{8}} + 4^{-(Nk-k-2)}\sigma^2$$

$$\leq 2^{-2kN} \sum_{l=1}^{kN-1} 2^{2(kN-l)+3} 2^{-\frac{(k-l/N)^2}{8}} + 4^{-(Nk-k-2)}\sigma^2$$

$$= 2^{-2kN} \cdot 2^3 \cdot 2^{8N^2} \sum_{l=1}^{kN-1} 2^{-\frac{(k-l/N-8N)^2}{8}} + 4^{-(Nk-k-2)}\sigma^2$$

$$< 2^{-2kN} \cdot 2^3 \cdot 2^{8N^2} \sum_{l=-\infty}^{\infty} 2^{-\frac{(k-l/N-8N)^2}{8}} + 4^{-(Nk-k-2)}\sigma^2$$

$$= 2^{-2kN} \cdot 2^3 \cdot 2^{8N^2} \sum_{l'=-\infty}^{\infty} 2^{-\frac{(l'/N)^2}{8}} + 4^{-(Nk-k-2)}\sigma^2$$

$$\leq 2^{-2kN} \cdot c_1 + 4^{-(Nk-k-2)}\sigma^2$$

$$\leq c\sigma^{2N}. \tag{59}$$

∎

It is worth noting that in the proposed family of codes, for each code, the asymptotic slope of the SDR curve is 1 (as we expected from the fact that for each code, the mapping is piecewise differentiable). We can mix the idea of this scheme with Scheme II of the previous section, to construct a family of mappings where for each of them, the asymptotic slope is $N$, and together, they maintain a bounded gap with the optimal SDR (in terms of dB):

**Type II - Family of robust mappings:** For $x + \frac{1}{2} = \left(\overline{0 \cdot b_1 b_2 b_3 ...}\right)_2$, we construct $f_k(x) = (s_1, s_2, ..., s_N)$ as

$$s_1 = \sum_{i=1}^{k}(2^{-i} + 2^{-k}(k-i))b_{(i-1)N+1} + 2^{-k-1}\left(\overline{0 \cdot b_{kN+1} 0 b_{kN+\frac{N(N+1)}{2}+1} b_{kN+\frac{N(N+1)}{2}+2}\cdots}\right)_2$$

$$s_2 = \sum_{i=1}^{k}(2^{-i} + 2^{-k}(k-i))b_{(i-1)N+2} + 2^{-k-1}\left(\overline{0 \cdot b_{kN+2} b_{kN+3} 0 b_{kN+\frac{(N+1)(N+2)}{2}+1}\cdots}\right)_2$$

$$...$$

$$s_N = \sum_{i=1}^{k}(2^{-i} + 2^{-k}(k-i))b_{(i-1)N+N} + 2^{-k-1}\left(\overline{0 \cdot b_{kN+\frac{N(N-1)}{2}+1} b_{kN+\frac{N(N-1)}{2}+2}\cdots}\right)_2$$

**Theorem 9** *In the proposed family of mappings (Type II), there are constants $c, c_1, c_2$, independent of $\sigma$ and $k$ (are only dependent on $N$) such that for every integer $k > 0$, if we use the modulation map $f_k(x)$,*

*i) For $2^{-k-1} < \sigma \leq 2^{-k}$,*

$$D \leq c\sigma^{2N}. \tag{60}$$

*ii) for any $\sigma < 2^{-k-1}$,*

$$D \leq c_1 \sigma^{2N} 2^{c_2 \sqrt{-\log_2 \sigma}}. \tag{61}$$

*Proof:* i) The probability that the first error occurs in the $l$th bit ($l = (i-1)N + j < kN$) of $x$ is bounded by $P_l \geq 2Q\left(\frac{k-i}{2}\right)$ and it results in an output squared error of at most $4^{-l+1}$, and when there is no error in the first $Nk$ bits, the squared error is $D' \leq 4^{-Nk}$. Therefore, by considering the union-bound over all possible errors, we have

$$D \leq \sum_{l=1}^{Nk} D_l \cdot P_l + D'$$

$$\leq \sum_{l=1}^{Nk} 4^{-l+1} \cdot 2Q\left(\frac{k}{2} - \frac{l}{2N}\right) + 4^{-Nk}$$

Similar to the proof of theorem 8, by using $Q(x) < e^{-\frac{x^2}{2}}$ and $2^{-k-1} < \sigma \leq 2^{-k}$, we have

$$D \leq \sum_{l=1}^{Nk-1} 4^{-l+1} e^{-\frac{(k-l/N)^2}{8}} + \sigma^{2N}$$

$$\leq c_4 4^{-kN} + \sigma^{2N}$$

$$\leq c\sigma^{2N}.$$

ii) Consider $z_i$ as the Gaussian noise on the $i$th channel and assume that $n$ is selected such that

$$k + \sum_{l=1}^{n+1} (lN + 1) \leq -\log_2 \sigma < k + \sum_{l=1}^{n+2} (lN + 1) \tag{62}$$

The probability that $|z_i| \geq 2^{-k-1-\sum_{l=1}^{n}(lN+1)}$ is negligible (it is bounded by $2Q\left(2^{(n+1)N}\right)$).

On the other hand, when $|z_i| < 2^{-k-1-\sum_{l=1}^{n}(lN+1)}$, the first $k + \sum_{l=0}^{n-1}(lN + i + 1)$ bits of $s_i$ can be decoded error-freely ($1 \leq i \leq N$) which include $k + \sum_{l=0}^{n-1}(lN + i)$ bits of $x$. Thus, the first $kN + \sum_{j=1}^{nN} j$ bits of $x$ can be decoded error-freely. Now, similar to the proof of theorem 6,

$$kN + \sum_{j=1}^{nN} j \geq \tag{63}$$

$$N\left(k + \sum_{l=1}^{n+2}(lN+1)\right) - c_5\sqrt{\sum_{l=1}^{n+1}(lN+1)} \tag{64}$$

$$N\left(k + \sum_{l=1}^{n+2}(lN+1)\right) - c_6\sqrt{k + \sum_{l=1}^{n+1}(lN+1)} \tag{65}$$

Therefore, by using the assumption (62),

$$kN + \sum_{j=1}^{nN} j \geq \tag{66}$$

$$- N \log_2 \sigma - c_6 \sqrt{-\log_2 \sigma} \qquad (67)$$

Therefore, the output distortion is bounded by

$$D \leq 2^{-2\left(kN + \sum_{j=1}^{nN} j\right)} + 2Q\left(2^{(n+1)N}\right) \qquad (68)$$

$$\leq 2^{2N \log_2 \sigma + 2c_6 \sqrt{-\log_2 \sigma}} + 2Q\left(2^{(n+1)N}\right) \qquad (69)$$

$$\implies D \leq c_1 \sigma^{2N} 2^{c_2 \sqrt{-\log_2 \sigma}}. \qquad (70)$$

∎

## VII. SIMULATION RESULTS

In figure 3, for a bandwidth expansion factor of 4, the performance of Scheme I (with parameters $\alpha = 3$ and 4) is compared with the shift-map scheme with $a = 3$. As we expect, for the shift-map scheme, the SDR curve saturates at slope 1, while the new scheme offers asymptotic slopes higher than one. For the proposed scheme, with parameters $\alpha_1 = 3$ and $\alpha_2 = 4$, the asymptotic slope is respectively $\beta_1 = \frac{4 \log 2}{\log 3}$ and $\beta_2 = \frac{4 \log 2}{\log 4} = 2$ (as expected from Theorem 3). Also, we see that the proposed scheme provides a graceful degradation in the low SNR region.

Figure 4 shows the performance of Scheme II for $N = 4$ dimensions. As it is shown in the figure, the asymptotic exponent of the SDR is close to the optimum value of 4, i.e. the bandwidth expansion ratio. The fluctuations of the slope of the curve is due to the fact that groups of consequent bits are assigned to each dimension, and for different ranges of SNR, errors in different dimensions become dominant (for example, for SNR values around 40-50dB, the error in the second layer of bits of $s_1$ becomes dominant in the overall squared error). By modifying Scheme II and assigning groups of bits of length $l' = i + k(N - 1)$ (instead of $l = i + kN$) to the $i$th dimension, we can slightly improve the performance in the middle SNR range. Asymptotic exponents of the SDR in both variations of Scheme II are the same.

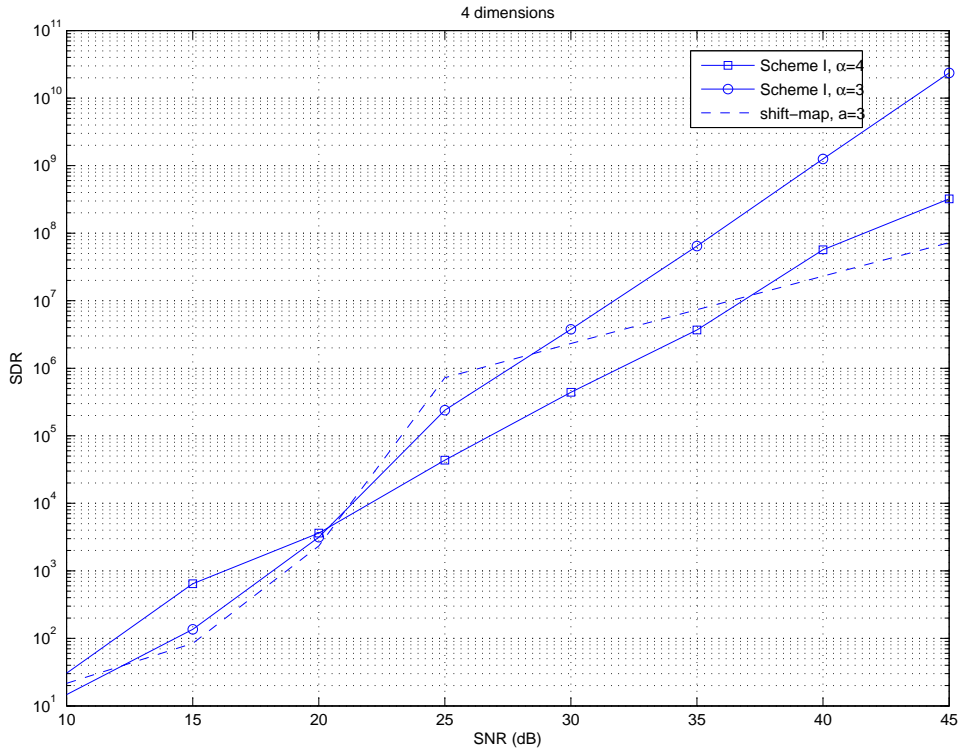

Fig. 3. The output SNR (or SDR) for the first proposed scheme (with $\alpha = 4$ and 3) and the shift-map scheme with $a = 3$. The bandwidth expansion is $N = 4$.

## VIII. Conclusions

To avoid the mild saturation effect in analog transmission (i.e. achieving the optimum scaling of the output distortion), one needs to use non-differentiable mappings (more precisely, mappings which are not differentiable on any interval). Two non-differentiable schemes are introduced in this paper. Both these schemes, which are minimum-delay schemes, outperform the traditional minimum-delay analog schemes, in terms of scaling of the output SDR. Also, one of them (Scheme II) achieves the optimum SDR scaling with a simple mapping (it achieves the asymptotic exponent $N$ for the SDR, versus SNR).

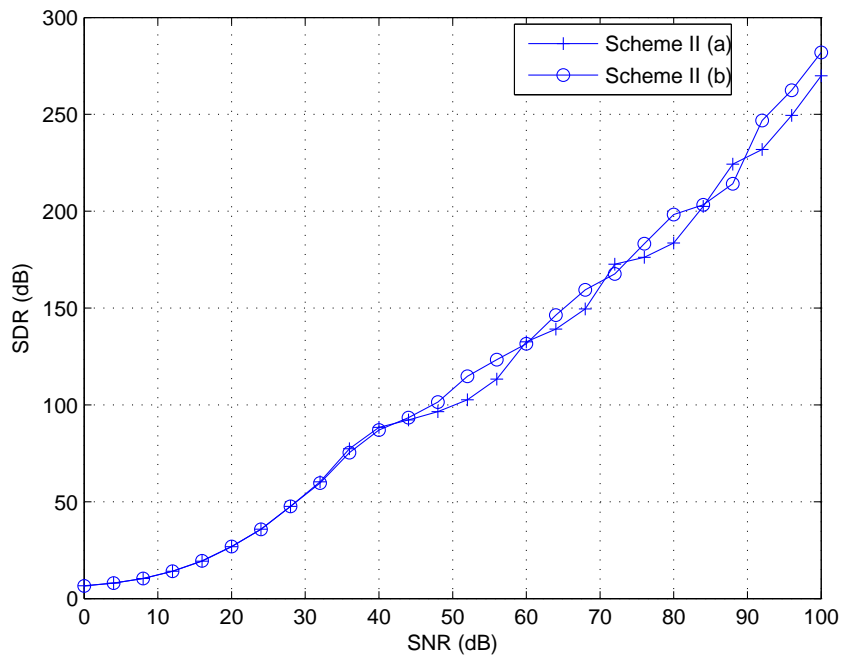

Fig. 4. Performance of Scheme II for $N = 4$ dimensions. (a) corresponds to the scheme introduced in Section V and (b) corresponds to the other variation of Scheme II, when groups of $l' = i + k(N-1)$ bits are considered.

## APPENDIX A: PROOF OF THEOREM 1

The set of modulated signals consists of $a^{N-1}$ parallel segments where the projection of each of them on the $i$th dimension has the length $a^{-(i-1)}$, hence, each segment has the length $\sqrt{1 + a^{-2} + ... + a^{-2(N-1)}}$. By considering the distance of their intersections with the hyperspace orthogonal to the $N$th dimension (which is at least $a^{-1}$) and the angular factor of these segments, respecting to the $s_N$-axis, because $a \geq 2$, we can bound the distance between two parallel segments of the modulated signal set as (see Fig. 1)

$$d \geq \frac{a^{-1}}{\sqrt{1 + a^{-2} + ... + a^{-2(N-1)}}} \geq \frac{a^{-1}}{\sqrt{1 + 2^{-2} + ... + 2^{-2(N-1)}}} \geq \frac{a^{-1}}{2} \qquad (71)$$

First, we consider the case of $\sigma\sqrt{-\log \sigma} \leq \frac{1}{16\sqrt{N}}$. Consider $a = \left\lfloor \frac{1}{8\sqrt{N}\sigma\sqrt{-\log \sigma}} \right\rfloor$. Probability of a jump to a wrong segment (during the decoding) is bounded by

25$$\Pr(jump) \leq 2Q\left(\frac{d}{2\sigma}\right) \leq 2Q\left(\frac{a^{-1}}{4\sigma}\right) \tag{72}$$

$$\leq 2Q\left(\frac{8\sqrt{N}\sigma\sqrt{-\log\sigma}}{4\sigma}\right). \tag{73}$$

By using $Q(x) \leq \frac{1}{2}e^{-\frac{x^2}{2}}$,

$$\Pr(jump) \leq e^{\frac{-(2\sqrt{N}\sqrt{-\log\sigma})^2}{2}} = e^{2N\log\sigma} = \sigma^{2N}. \tag{74}$$

On the other hand, each segment of the modulated signal set is a segment of the source signal set, stretched by a factor of $a^{N-1}\sqrt{1+a^{-2}+...+a^{-2(N-1)}}$ (its length is changed from $\frac{1}{a^{N-1}}$ to $\sqrt{1+a^{-2}+...+a^{-2(N-1)}}$). Therefore, assuming the correct segment decoding, the average distortion is the variance of the channel noise divided by $\left(a^{N-1}\sqrt{1+a^{-2}+...+a^{-2(N-1)}}\right)^2$:

$$\mathrm{E}\left\{|\tilde{x}-x|^2|no\ jump\right\} = \tag{75}$$

$$\frac{\sigma^2}{\left(a^{N-1}\sqrt{1+a^{-2}+...+a^{-2(N-1)}}\right)^2} \leq \tag{76}$$

$$\frac{\sigma^2}{a^{2(N-1)}} = \frac{\sigma^2}{\left[\frac{1}{8\sqrt{N}\sigma\sqrt{-\log\sigma}}\right]^{2(N-1)}} \leq c_1 \sigma^{2N}(-\log\sigma)^{N-1} \tag{77}$$

where $\tilde{x}$ is the estimate of $x$ and $c_1$ is independent of $a$ and $\sigma$ and only depends on $N$. Now, because $\mathrm{E}\left\{|\tilde{x}-x|^2|jump\right\}$ and $\Pr(no\ jump)$ are bounded by 1,

$$D = \Pr(jump) \cdot \mathrm{E}\left\{|\tilde{x}-x|^2|jump\right\} + \Pr(no\ jump) \cdot \mathrm{E}\left\{|\tilde{x}-x|^2|no\ jump\right\} \tag{78}$$

$$\Longrightarrow D \leq \Pr(jump) + \mathrm{E}\left\{|\tilde{x}-x|^2|no\ jump\right\} \tag{79}$$

$$\leq c_2 \sigma^{2N}(-\log\sigma)^{N-1}, \quad \text{for } \sigma\sqrt{-\log\sigma} \leq \frac{1}{16\sqrt{N}}. \tag{80}$$

On the other hand, for $\sigma\sqrt{-\log\sigma} > \frac{1}{16\sqrt{N}}$,

$$D \leq 1 = \sigma^{-2N}(-\log\sigma)^{-(N-1)} \cdot \sigma^{2N}(-\log\sigma)^{N-1} \tag{81}$$

$$= \left(\sigma\sqrt{-\log\sigma}\right)^{-2N} \cdot (-\log\sigma) \cdot \sigma^{2N}(-\log\sigma)^{N-1} \tag{82}$$

$$< \left(\frac{1}{16\sqrt{N}}\right)^{-2N} \cdot (-\log\sqrt{2}) \cdot \sigma^{2N}(-\log\sigma)^{N-1} \tag{83}$$

$$\leq c_3 \sigma^{2N}(-\log\sigma)^{N-1}. \tag{84}$$

Therefore, by combining these two bounds together, we obtain

$$D \leq c\sigma^{2N}(-\log\sigma)^{N-1}. \tag{85}$$

### APPENDIX B: PROOF OF THEOREM 2

We consider two cases:

Case 1) $a \leq \frac{4}{\sigma\sqrt{-\log\sigma}}$:

Each segment of the modulated signal set is a segment of the source signal set, scaled by a factor of $a^{N-1}\sqrt{1 + a^{-2} + ... + a^{-2(N-1)}}$, hence

$$D \geq \mathrm{E}\left\{|\tilde{x} - x|^2 | no\ jump\right\} \tag{86}$$

$$= \frac{\sigma^2}{\left(a^{N-1}\sqrt{1 + a^{-2} + ... + a^{-2(N-1)}}\right)^2} \tag{87}$$

$$\geq \frac{\sigma^2}{2a^{2(N-1)}} \tag{88}$$

$$\geq c_1 \sigma^{2N}(-\log\sigma)^{N-1} \tag{89}$$

Case 2) $\frac{2^{l+1}}{\sigma\sqrt{-\log\sigma}} < a \leq \frac{2^{l+2}}{\sigma\sqrt{-\log\sigma}}$, for $l \geq 1$:

In this case, we bound the output distortion by the average distortion caused by a large jump to another segment. Let $z_1$ be the additive noise in the first dimension and $f(x) = \mathbf{s}$ the modulated vector corresponding to the source sample $x$.

For any point in the interval $-\frac{1}{2} + (k-1)a^{-1} < x \leq -\frac{1}{2} + ka^{-1}$ (for $1 \leq k \leq a - 2^{l+1}$), when $z_1 > 2^{l+1}a^{-1}$, for any point $x' \leq x + 2^l a^{-1}$, the received point $f(x) + \mathbf{z}$ is closer to $f\left(x' + 2^l a^{-1}\right)$ than $f(x)$. Therefore, the decoded signal is $\tilde{x} > x + 2^l a^{-1}$. Thus, in this case, the squared error is at least $\left(2^l a^{-1}\right)^2$. Therefore, the average distortion is lower bounded by

$$D \geq \Pr\left\{-\frac{1}{2} < x \leq \frac{1}{2} - 2^{l+1}a^{-1}\right\} \cdot \Pr\left\{z_1 > 2^{l+1}a^{-1}\right\} \cdot \left(2^l a^{-1}\right)^2 \tag{90}$$

$$= \left(1 - 2^{l+1}a^{-1}\right) \cdot Q\left(\frac{2^{l+1}a^{-1}}{\sigma}\right) \cdot \left(2^l a^{-1}\right)^2 \tag{91}$$

$$\geq \left(1 - \sigma\sqrt{-\log\sigma}\right) \cdot Q\left(\frac{\sigma\sqrt{-\log\sigma}}{\sigma}\right) \cdot \left(\frac{\sigma\sqrt{-\log\sigma}}{2^2}\right)^2 \tag{92}$$

$$= \left(1 - \sigma\sqrt{-\log\sigma}\right) \cdot Q\left(\sqrt{-\log\sigma}\right) \cdot \frac{\sigma^2(-\log\sigma)}{2^4} \tag{93}$$

By using $e^{-x^2} < Q(x)$,

$$D \geq \left(1 - \sigma\sqrt{-\log\sigma}\right) \cdot \sigma \cdot \frac{\sigma^2(-\log\sigma)}{2^4} \tag{94}$$

$$\implies D \geq c_2 \sigma^3 (-\log\sigma). \tag{95}$$

By combining the bounds (for two cases), and noting that $\sigma^2 \leq \frac{1}{2}$,

$$D \geq \min\left\{c_2 \sigma^3 (-\log\sigma), c_1 \sigma^{2N}(-\log\sigma)^{N-1}\right\} \tag{96}$$

$$D \geq c' \sigma^{2N}(-\log\sigma)^{N-1} \quad \text{for } N \geq 2. \tag{97}$$

APPENDIX C: CODING FOR UNBOUNDED SOURCES

Consider $\{X_i\}_{i=1}^{\infty}$ as an arbitrary memoryless i.i.d source. We show that the results of Section V can be extended for non-uniform sources, to construct robust joint-source channel codes with a constraint on the average power. Without loss of generality, we can assume the variance of the source to be equal to 1. For the source sample $x$, we can write it as $x = x_1 + x_2$ where $x_1$ is an integer, $-\frac{1}{2} \leq x_2 < \frac{1}{2}$, and $x_2 + \frac{1}{2} = \left(\overline{0 \cdot b_1 b_2 b_3 ...}\right)_2$. Now, we construct the $N$-dimensional transmission vector as $\mathbf{s}' = (s_1', s_2', ..., s_N') = \left(x_1 + s_1 - \frac{1}{2}, s_2 - \frac{1}{2}, ..., s_N - \frac{1}{2}\right)$, where $s_1, ..., s_N$ are constructed using (36) in section V. Let $D_1$ be the distortion conditioned on correct decoding of $x_1$. Similar to the proof Theorem 6, we can show that the $D_1$ is upper bounded by

$$D_1 \leq c_1 \sigma^{2N} 2^{c_2 \sqrt{-\log \sigma}} \tag{98}$$

where $c_1$ and $c_2$ depend only on $N$.

Now, we bound the distortion $D_2$, for the case that $x_1$ is not decoded correctly. Since $s_1$ is constructed by scheme II (in Section V), $s_1$ is between 0 and $(\overline{0.10111\cdots})_2$, hence $0 \leq s_1 < \frac{3}{4}$. To have an error of $|x_1 - \widetilde{x}_1| = k$, the amplitude of the noise on the first dimension should be greater than $\frac{k-\frac{3}{4}}{2}$, hence its probability is bounded by $2Q\left(\frac{k-\frac{3}{4}}{2\sigma}\right)$. When $|x_1 - \widetilde{x}_1| = k$, the overall squared error is lower bounded by

$$|x - \widetilde{x}| \leq |x_1 - \widetilde{x}_1| + |x_2 - \widetilde{x}_2| \leq k + 1. \tag{99}$$

Therefore, by using the union bound for all values of $k$, the distortion $D_2$ is lower bounded by

$$D_2 \leq \sum_{k=1}^{\infty} 2Q\left(\frac{k-\frac{3}{4}}{2\sigma}\right)(k+1) \tag{100}$$

$$\leq \sum_{k=1}^{\infty} e^{-\frac{\left(\frac{k-\frac{3}{4}}{2\sigma}\right)^2}{2}} \cdot (k+1) \tag{101}$$

$$\leq c_3 e^{\frac{-1}{128\sigma^2}}. \tag{102}$$

Thus, $D \leq D_1 + D_2 \leq c_4 \sigma^{2N} 2^{c_2 \sqrt{-\log \sigma}}$.

To finish the proof, we only need to show that the average transmitted power is bounded. For $s'_2, ..., s'_N$, the transmitted power is bounded as $|s'_i|^2 \leq \frac{1}{4}$. For $s'_1$,

$$|s'_1|^2 = \left|x_1 + s_1 - \frac{1}{2}\right|^2 \leq \left(|x_1| + \left|s_1 - \frac{1}{2}\right|\right)^2 \tag{103}$$

$$\leq \left(|x| + \frac{1}{2} + \frac{1}{2}\right)^2 = (|x| + 1)^2 \tag{104}$$

Thus, using the Cauchy-Schwarz inequality,

$$\mathrm{E}\left\{|s'_i|^2\right\} \leq \mathrm{E}\left\{(|x|+1)^2\right\} \leq \left(\sqrt{\mathrm{E}|x|^2} + 1\right)^2 \tag{105}$$

$$\leq (1+1)^2 = 4. \tag{106}$$

### APPENDIX D: PROOF OF THEOREM 7

We consider two cases for $a$, the scaling factor,

Case 1) $a \leq 2^{\frac{2(N-M)}{M}+4} \sigma^{-\frac{(N-M)}{M}} (-\log \sigma)^{\frac{-(N-M)}{2M}}$:

Each subset of the modulated signal set is the scaled version of a segment of the source signal set by a factor of $a$, hence, we can lower bound the distortion by only considering the case that the subset is decoded correctly and there is no jump to adjacent subsets,

$$D \geq \mathrm{E}\left\{|\tilde{x} - x|^2 | no\ jump\right\} \tag{107}$$

$$= \frac{\sigma^2}{a^2} \tag{108}$$

$$\geq c_4 \sigma^{\frac{2N}{M}} (-\log \sigma)^{\frac{N-M}{M}} \tag{109}$$

Case 2) $2^{l+1+\frac{2(N-M)}{M}} < \frac{a}{\sigma^{-\frac{(N-M)}{M}}(-\log \sigma)^{\frac{-(N-M)}{2M}}} \leq 2^{l+2+\frac{2(N-M)}{M}}$ for $l \geq 3$:

In this case, we bound the output distortion by the average distortion caused by a jump to another subset. Without loss of generality[7], we can consider $\sigma < \left(\frac{1}{e}\right)$, hence $2^{-l}a > 8$. First, we show that there are two constants $c_5$ and $c_6$ (independent of $a$ and $\sigma$) such that probability of an squared error of at least $c_5 \left(2^{-l}a\right)^{-2}$ is lower bounded by

$$\Pr(jump) \geq c_6 Q\left(\sqrt{-\log \sigma}\right) \geq c_6 \sigma \tag{110}$$

---

[7]For $1 < \sigma < \frac{1}{e}$, the distortion $D$ is larger than $D_{\frac{1}{e}}$ (the distortion for $\sigma = \frac{1}{e}$), hence $D \geq D_{\frac{1}{e}} > D_{\frac{1}{e}} \sigma^{\frac{2N}{M}}(-\log \sigma)^{\frac{N-M}{M}}$, and $D_{\frac{1}{e}}$ depends only on $N$.

By considering the power constraint, the maximum distance of each source sample to its quantization point is upper bounded by

$$d_{max} \leq \frac{1}{a}. \tag{111}$$

We can partition the $M$-dimensional uniform source to $n = \left(\lfloor \frac{a}{2^l} \rfloor\right)^M \geq \left(\frac{a}{2^{l+1}}\right)^M$ cubes of size $s = \frac{1}{\lfloor \frac{a}{2^l} \rfloor} \geq \frac{2^l}{a} \geq 2^l d_{max}$. We consider $\mathcal{B}_i$ as the union of the quantization regions whose center is in the $i$th cube ($1 \leq i \leq n$). Because the decoding of digital and analog parts are done separately, the $(N-M)$-dimensional subspace (dedicated to send the quantization points) can be partitioned to $n$ decoding subsets, corresponding to regions $\mathcal{B}_1, ..., \mathcal{B}_n$. If we consider $\mathcal{C}_1, ..., \mathcal{C}_n$, the intersections of these decoding regions and the $(N-M)$-dimensional cube of size 4, centered at the origin, at least $\frac{n}{2}$ of them have volume less than $2\left(\frac{(4)^{N-M}}{n}\right) \leq \frac{(4)^{N-M}}{\left(2^{-l-1}a\right)^M} \leq 2\sigma^{N-M}(-\log \sigma)^{\frac{N-M}{2}}$. This volume is less than the volume of an $(N-M)$-dimensional sphere of radius $\sigma(-\log \sigma)^{\frac{1}{2}}$. Thus, for any point inside $\mathcal{B}_i$ with this property, the probability of being decoded to a wrong subset $\mathcal{B}_j$ is at least equal to the probability that the amplitude of the noise is larger than the radius of that sphere (i.e. $\sigma(-\log \sigma)^{\frac{1}{2}}$). This probability is lower bounded by $\Pr\left\{z_1 > \sigma(-\log \sigma)^{\frac{1}{2}}\right\} = Q\left(\sqrt{-\log \sigma}\right) \geq \sigma$. Now, for the cubes corresponding to these subsets, we consider points inside a smaller cube of size $\frac{s}{2}$, with the same center.

For these points, at least with probability $\sigma$, decoder finds a wrong quantization region where the distance of its center and the original point is at least $\frac{s-\frac{s}{2}}{2} = \frac{s}{4} \geq \frac{2^{l-2}}{a}$, hence, the final squared error is at least $\left(\frac{2^{l-2}}{a} - d_{max}\right)^2 \geq \left(\frac{2^{l-2}}{a} - \frac{1}{a}\right)^2 \geq c_5 \left(2^{-l}a\right)^{-2}$.

Because at least half of the $n$ subsets have the mentioned property, the overall probability of having this kind of points as the source is at least $\frac{1}{2}2^{-M}$, and in transmitting these points, with a probability which is lower bounded by $\sigma$, the squared error is at least $c_5 \left(2^{-l}a\right)^{-2}$. Therefore, the distortion is lower bounded by

$$D \geq \frac{1}{2}2^{-M} \cdot \sigma \cdot c_5 \left(2^{-l}a\right)^{-2} \geq c_7 \sigma \left(2^{-l}a\right)^{-2}$$

$$\geq c_8 \sigma \cdot \sigma^{\frac{2(N-M)}{M}} (-\log \sigma)^{\frac{N-M}{M}}$$

$$= c_8 \sigma^{\frac{2N-M}{M}} (-\log \sigma)^{\frac{N-M}{M}}. \tag{112}$$

Finally, by considering the minimum of (109) and (112), we conclude

$$D \geq c\sigma^{\frac{2N}{M}} (-\log \sigma)^{\frac{N-M}{M}}. \tag{113}$$